\documentclass[twocolumn,prl,showpacs,preprintnumbers,amsmath,amssymb]{revtex4}
\usepackage{graphicx,epsf}
\setkeys{Gin}{width=8.6cm,keepaspectratio}
\usepackage{dcolumn}
\usepackage{bm}

\begin{document}

\title{Quasi-stationary criticality of the Order-Parameter of
the $d=3$ Random-Field Ising Antiferromagnet
$\rm Fe_{0.85}Zn_{0.15}F_2$: A Synchrotron X-ray Scattering Study}

\author{F. Ye$^1$,  L. Zhou$^2$, S. A. Meyer$^1$, L. J. Shelton$^1$,
D. P. Belanger$^1$, L. Lu$^3$, S. Larochelle$^4$, and M. Greven$^{2,3}$}

\affiliation{$^1$Department of Physics, University of California,
 Santa Cruz, California 95064}
\affiliation{$^2$Stanford Synchrotron Radiation Laboratory, Stanford,
 California 94309}
\affiliation{$^3$Department of Applied Physics, Stanford University, Stanford,
 California 94305}
\affiliation{$^4$Department of Physics, Stanford University, Stanford,
 California 94305}

\date{\today}

\begin{abstract}
The critical exponent $\beta =0.17 \pm 0.01$ for the three-dimensional
random-field Ising
model (RFIM) order parameter upon zero-field cooling (ZFC) has been determined
using extinction-free magnetic x-ray scattering techniques for
$\rm Fe_{0.85}Zn_{0.15}F_2$.  This result is consistent with
other exponents determined for the RFIM in that Rushbrooke
scaling is satisfied.  Nevertheless, there is poor agreement with
equilibrium computer simulations, and the ZFC results do not agree
with field-cooling (FC) results.  We present details of hysteresis in
Bragg scattering amplitudes and line shapes that help elucidate the
effects of thermal cycling in the RFIM, as realized in dilute
antiferromagnets in an applied field.  We show that the ZFC
critical-like behavior is consistent with a second-order phase transitions,
albeit quasi-stationary rather than truly equilibrium in nature, as
evident from the large thermal hysteresis observed near the transition.
\end{abstract}

\pacs{61.10.Nz, 75.40.Cx, 75.50.Ee, 75.50.Lk}

\maketitle

\section{INTRODUCTION}

One of the classic models of a phase transition
in an intrinsically disordered system is the
three-dimensional ($d=3$) random-field Ising model (RFIM).
After nearly three decades of progress in experiments, simulation
and theory, the phase transition in this system is
not yet well understood and remains a challenge to our
fundamental understanding of the statistical physics
of disordered systems.
Extensive Monte Carlo simulations and exact ground state
calculations have provided evidence
for equilibrium critical exponents, including
a value for the order-parameter exponent close to zero,
but with a specific heat exponent that
is not yet well determined~\cite{wm06,wm05,hy01,mf02,mf06}.
The specific heat exponent value, however, has been
determined in experiments in the dilute
antiferromagnet $\rm Fe_{\rm x}Zn_{\rm 1-x}F_2$ at various
magnetic concentrations $x$~\cite {sb98,ysmb04} to be very close to
zero upon warming through the phase transition.  This is indicated
by the apparent symmetric, logarithmic divergence as well as
the field scaling of the optical birefringence, Faraday rotation,
and specific heat amplitudes~\cite{pkb88}.
The experimental determination of the order parameter has
proven more elusive because the large number of vacancies at low magnetic
concentrations allows the system
to easily form domains that obscure the order-parameter
critical behavior.

An avenue for determining the order-parameter
exponent opened once it was understood that at higher
magnetic concentrations these domains do not form.  
In the best studied system, the dilute antiferromagnet
$\rm Fe_{\rm x}Zn_{\rm 1-x}F_2$ in a field applied along the easy axis,
domains do not form
for magnetic concentrations $x > 0.75$~\cite{bybf04,bb00a,gjd05}.
As we will discuss below, the experimental results nevertheless
do not agree well with the equilibrium Monte Carlo and exact ground state
calculations.  Indeed, the experimentally-observed transition
is clearly not in equilibrium, although critical behavior
can be determined under conditions of monotonically
increasing temperature.  In this study,
we describe the apparent critical behavior and characterize the
hysteresis observed upon crossing the phase boundary.
We also discuss temperature reversals just below the phase boundary.
In this way, we describe the unusual circumstance of
self-consistent critical-like behavior in a clearly nonequilibrium
system.  The discrepancy with simulations is a consequence
of the simulations being done under equilibrium conditions,
something apparently not realized in the macroscopic experimental system.

The phase transition in the $d=3$
RFIM system $ \rm Fe_{0.85}Zn_{0.15}F_2$ has been
characterized in great detail for the zero-field-cooling (ZFC)
procedure in which the sample is cooled in zero field,
the field is raised, and the sample is warmed in constant
field across the phase boundary.
The transition appears to be second-order under ZFC;
all of the critical exponents, measured
at very small reduced temperatures, are self-consistent
in that they appear to satisfy the Rushbrooke equation, as described
below.  Nevertheless, it is well known that
different behavior is observed upon field-cooling (FC),
in which the sample is cooled
across the phase boundary in the field.  This has been
observed for this sample in x-ray scattering~\cite{yzllbgl02},
neutron scattering~\cite{ymkybsfa03}, optical birefringence~\cite{ysmb04} and
optical Faraday rotation~\cite{ysmb04} experiments.  Hence, the ZFC
phase transition is extraordinary in that it appears to
be second-order if the temperature reversals are avoided
but, in light of the hysteresis, it
does not take place under equilibrium conditions as would
be the usual case.  To better characterize the apparent RFIM
transition, the FC critical behavior, and
the hysteresis in general, needs to be explored
more fully.  We have done this for the order
parameter in $ \rm Fe_{0.85}Zn_{0.15}F_2$ using magnetic
x-ray scattering.

Characterization of the order parameter, the staggered
magnetization ($M_s$), is achieved
by determining the temperature dependence of the
antiferromagnetic Bragg scattering intensity, which
is proportional to $M_s$,
versus the temperature $T$. The order parameter is expected to behave as
\begin{equation}
M_s=M_0t^{\beta}
\label{eqn:xray-order}
\end{equation}
for $t<<1$, where $t=(T_{\rm c}(H)-T)/T_{\rm c}(H)$ is the reduced temperature
and $T_{\rm c}(H)$ is the transition temperature.
As discussed in a previous Letter~\cite{yzllbgl02}, which focused
on the ZFC measurements, neutron scattering techniques can not
be used to characterize the order parameter critical behavior
in high-crystalline-quality bulk crystals
because of severe extinction effects, which tend to modify the temperature
dependence of the Bragg intensity as the temperature is lowered.
The effect arises when the scattering sample region
is so thick and the crystal is so perfect that the beam
is selectively depleted of
neutrons that satisfy the Bragg condition for scattering.
This prevents an accurate determination of the
exponent $\beta$.  The x-ray technique, on the other
hand, is essentially free of extinction effects
and the order-parameter criticality can be measured
accurately near $T_{\rm c}(H)$~\cite{yzllbgl02}.
This is possible since the thickness of the scattering
region is limited by the strong temperature-independent
charge scattering and, within this region, the
beam is never depleted of x-rays meeting the Bragg
condition since the magnetic scattering is weak.

A magnetic concentration near the one in this
study, $x=0.85$, is crucial
to this order parameter characterization.
Many prior attempts~\cite{hfbt93,bwshnlrl96}
to determine the critical behavior
of the order parameter proved unsuccessful because
$x<x_{\rm v}$, where $x_{\rm v} = 0.754$ is the magnetic vacancy
percolation~\cite{bb00a}.  Domain formation obscures
the RFIM critical behavior below the transition
at $T_{\rm c}(H)$ in $\rm Fe_{\rm x}Zn_{\rm 1-x}F_2$
and its less anisotropic ({\em i.e.}, smaller Ising anisotropy) isomorph
$\rm Mn_{\rm x}Zn_{\rm 2-x}F_2$ for $x<x_{\rm v}$.
The concentration $x=0.85$, while being greater than $x_{\rm v}$,
is low enough to generate
significant random-field effects at the maximum field of our
experiment, $H = 11$~T.
It is, at the same time, high enough to avoid the complication
of contributions to the scattering intensity from
fractal-like percolating vacancy
structures~\cite{bybf04} that appear close to $x_{\rm v}$.

\begin{figure}[t!]
 \centerline{
  \epsfxsize=3.8in
   \epsfbox{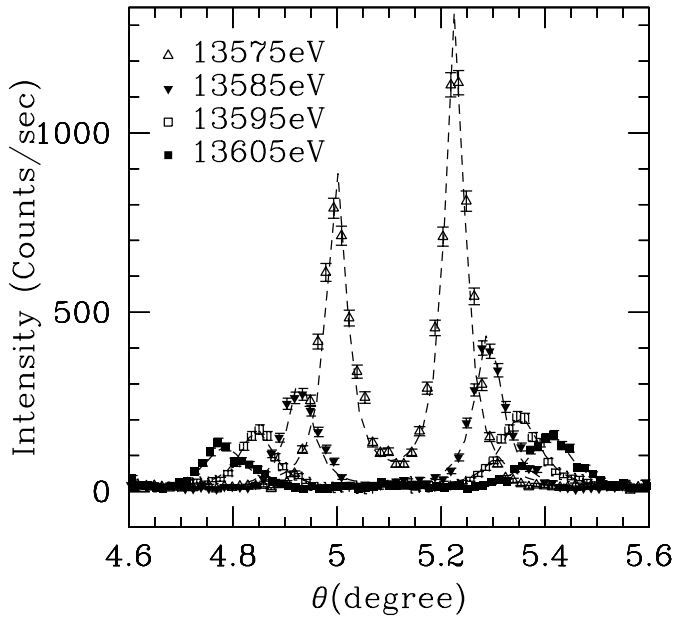}
}
\caption[Multiple Scattering vs.\ x-ray Energy]
{The energy of the x-ray was carefully fine-tuned to ensure that the
effect of multiple scattering is minimized
near the $(1 \ 0 \ 0)$ antiferromagnetic Bragg point, which
occurs near $\theta = 5.1$ degrees.  The temperature is several
degrees above the transition temperature, so no magnetic
peak is present.  The multiple scattering peaks appear
symmetrically on either side of the Bragg point.
These data demonstrate that the shoulder peaks move apart
and become weaker as the energy is tuned close to $13605$~eV.
The curves are Lorentzian fits.  The widths increase
as the peaks move apart and decrease in amplitude,
allowing the Bragg peak to be clearly discernible at low
temperatures, as shown in Fig.\ 2.
\label{fig:xray.mulscatter}
}
\end{figure}

\section{EXPERIMENTAL DETAILS}

\begin{figure}[t!]
 \centerline{
  \epsfxsize=3.8in
   \epsfbox{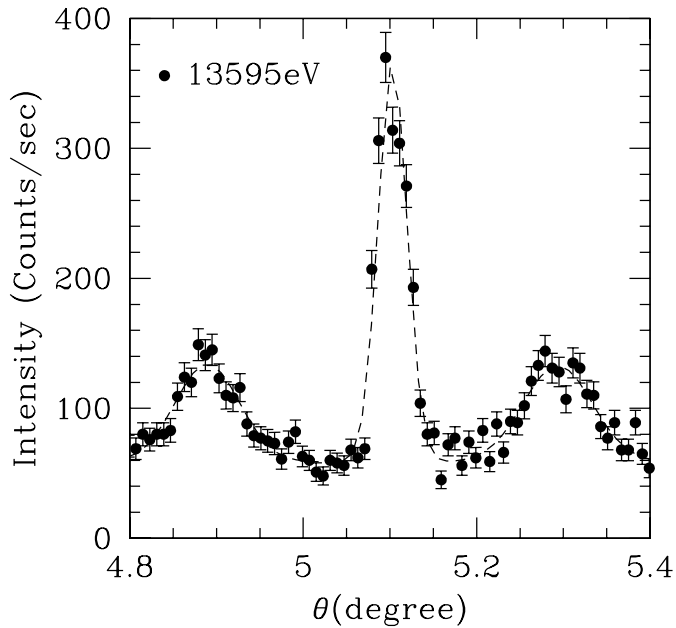}
 }
\caption[Typical Scan at Low T]
{Typical scattering at low temperature for $H=0$.
The energy is adjusted to minimize the multiple scattering,
which is nevertheless still present, but well separated
from the Bragg peak.  The magnetic Bragg peak is about
half as wide as the multiscattering peaks, as determined from
the Lorentzian fits shown in the figure.
\label{fig:xray.goodscatter}
}
\end{figure}

We are interested in the magnetic scattering intensity near the
antiferromagnetic zone center $(1 \ 0 \ 0)$.  Hence, we need not worry
about strong
charge scattering contributions and can easily discern the
relatively weak magnetic response.
The measurements were made at the high-field magnet facility
on beam line 7-2 of the Stanford Synchrotron Radiation Laboratory.
A monochromatic x-ray beam was obtained using a Si(111)
double-crystal monochromator from a spectrum produced
by a wiggler insertion device.  The x-ray energy was carefully
tuned to a value between 13.5 and 14 keV to minimize the effect of
energy-sensitive multiple scattering peaks around the
magnetic Bragg point~\cite{hfbt93}.  The incident x-ray
energy was well-defined, within
about 10 eV, whereas the detector had
a half-width energy resolution of 300 eV.
As shown in Fig.\ \ref{fig:xray.mulscatter}, at temperatures above
the transition temperature, where no magnetic scattering occurs,
the x-ray energy was adjusted until
the multiple scattering peaks move apart as far as possible~\cite{hfbt93}.
From Fig.\ \ref{fig:xray.goodscatter} it is clear that the multiple
scattering peaks do not affect the analysis
of the magnetic peak below the transition for an x-ray energy
of $13595$~eV.

The sample has a finely polished face, approximately
$13$~mm$^2$ in area, and is $0.9$~mm thick,
with the $a$-axis perpendicular
to the polished face.  It was mounted such that
the $c$-axis was along the vertical
field. The well polished surface prevented spurious effects
such as those observed in previous experiments~\cite{hfbt93},
where a transition-like behavior was observed that disappeared
when that the faces were polished.  Presumably, this is a result
of strong pinning at the locations of the scratches which
prevented RFIM correlations to develop.
The chemical homogeneity of the crystal was determined
by a room temperature optical linear
birefringence technique~\cite{kfjb88}
to be $0.45\%$/cm. The rounding of the transition is approximately
$50$~mK, corresponding to a tiny reduced temperature
of $0.0004$ on either side of the transition.
The crystal was mounted on a thin silicon bar of dimensions
$0.8 \times 1.5 \times 15 \rm {mm}^3$ and placed
in an atmosphere of approximately
10-20 mbar of Helium gas to achieve stabilization
of the sample temperature to within approximately 10 mK.
The zero-field transition temperature was measured to
be $T_{\rm N}=66.7$~K, consistent with birefringence measurements
on the same sample~\cite{ysmb04} and with a magnetic concentration
$x=0.85$~\cite{bkbj80}.

The lattice parameters of the sample were determined to be
approximately $a=4.68$~\AA\ and $c=3.27$~\AA\ near the transition
temperature. The full-widths-at-half-maximum measured at the
(1 0 0) magnetic Bragg point were $4 \times 10^{-4}$, $4 \times 10^{-3}$,
and $4 \times 10^{-3}$ reciprocal lattice units (r.l.u.)
for the transverse, longitudinal and vertical
directions, respectively.
Three conventional thermal-cycling procedures were
employed: ZFC; FC; and field-heating (FH).
As described previously, in ZFC the sample is cooled
across the phase boundary with $H=0$
and data are taken while warming with $H>0$, whereas
in FC the data are taken while cooling across the
phase boundary with $H>0$. 
In the FH procedure, the sample is first cooled through
the transition in a field, and then the data are taken
at that field value at successively higher temperatures.
During each procedure, when the sample temperature was not
continuously changed, the sample was
held at each temperature for at least
20 min before taking data to ensure that the
temperature and system stabilized.  The temperature
was stabilized before $q$ scans were obtained.  However,
when monitoring the peak intensity, it was more convenient
to let the temperature change continuously.
The data taken at $H=0$ and $H=11$~T in this study, except where noted,
were obtained in
transverse $(1 \ q \ 0)$ scans typically consisting
of 41 points, about 15 of which covered the Bragg peak.
At each point, the intensity was counted for 30 to 45 seconds,
depending on the temperature of the scan.  At other fields, Bragg
intensities were obtained at $q=0$ only.

\begin{figure}[t!]
 \centerline{
  \epsfxsize=3.8in
   \epsfbox{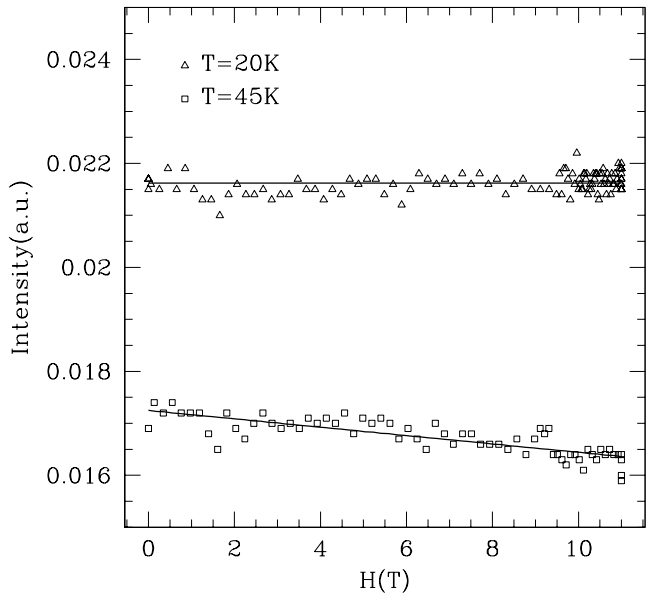}
 }
\caption[Field dependence of Bragg Peak Intensity]
{ Field dependence of the Bragg peak intensity measured at $T=20$~K
and $45$~K after cooling in $H=0$. The long-range order remains stable in
fields up to $H=11$~T, {\em i.e.}, no apparent phase boundaries are
crossed.
\label{fig:fieldscan_lowt}
}
\end{figure}

The antiferromagnetic transition for $x=0.85$ has been
shown to be stable in applied fields as high as $H=18$~T~\cite{samb02}.
To make sure that the ordered system is well-behaved
at low $T$ as we raised the field, we first cooled the
sample in zero field to $20$~K.
The magnetic field was then slowly raised
to $11$~T at a rate of 0.4 T/min for $H \le 9$~T and
0.1 T/min for $9<H\le 11$~T. The peak intensity of the order parameter
was monitored as we raised
the field.  Another set of measurements was similarly taken at $T=45$~K.
The results at both temperatures are shown
in Fig.\ \ref{fig:fieldscan_lowt}.
The intensity at $T=20$~K remains essentially unchanged as the
field is increased since the
data are all taken deep within the ordered region.  The result for
$45$~K shows a slight field dependence since, for this
case, the phase boundary (at $63.7$~K for $H=11$~T)
is approached more closely, although it is
still quite far away.  Those data demonstrate the
stability of the order, {\em i.e.}, no apparent phase boundaries were
crossed.  No change in the line width
of the Bragg scattering was observed upon application of the field,
further attesting to the stability of the order at low temperatures.

\begin{figure}[t!]
 \centerline{
  \epsfxsize=3.8in
   \epsfbox{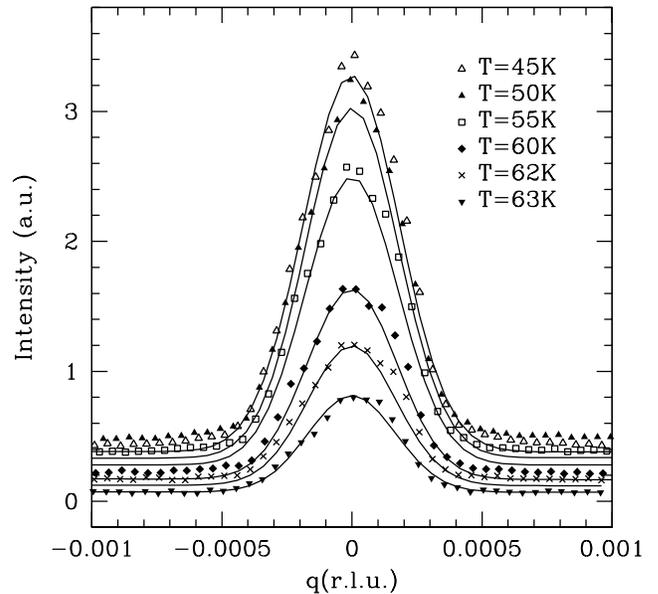}
 }
\caption[Transverse Scans at difference Temperatures at $H=11$~T]
{Representative transverse scans for the different temperatures
below $T_{\rm c}(H) = 63.7$~K taken with
$H=11$~T after cooling in $H=0$. Each scan is displaced vertically by 0.1 units
from the scan below it for clarity.  The solid curves
are results of least-squares fits to
a Gaussian line shapes with a half-width-at-half-maximum
equal to $2.1\times10^{-4}$ r.l.u.
}
\label{fig:discretescan}
\end{figure}

The x-ray scattering technique has the advantage of
very high momentum resolution, allowing a detailed study of the
Bragg peak line shapes.  It is important to use fine
collimating slit widths to observe details of the
scattering line shapes.  For this purpose,
the horizontal slits were configured
to be $0.5$~mm, approximately $30$~cm in front
of the sample and $1.1$~mm the same distance
after the sample.  At approximately $70$~cm behind
the sample is a second slit which was set at $1.35$~mm.
Care was taken to ensure that the sample
was well aligned for all measurements while using this
narrow slit configuration.  For measurements of the
peak intensity versus temperature, the slits after the
sample were opened slightly to reduce the sensitivity
of the intensity to the precise alignment of the sample and
beam so that wide ranges of $T$ and $H$ could be accessed without
continual adjustments.

Figure \ref{fig:discretescan} shows typical transverse
scans at different temperatures in a field $H=11$~T after ZFC.
The peak line shapes appear to be consistent with resolution-limited Gaussians
for all temperatures below the transition temperature
for $H=11$~T, $T_{\rm c}=63.7$~K.
As the temperature approaches $T_{\rm c}$, the intensity
gradually diminishes.  The relatively very weak critical
scattering is not apparent in this figure.

\begin{figure}[t!]
 \centerline{
  \epsfxsize=3.8in
   \epsfbox{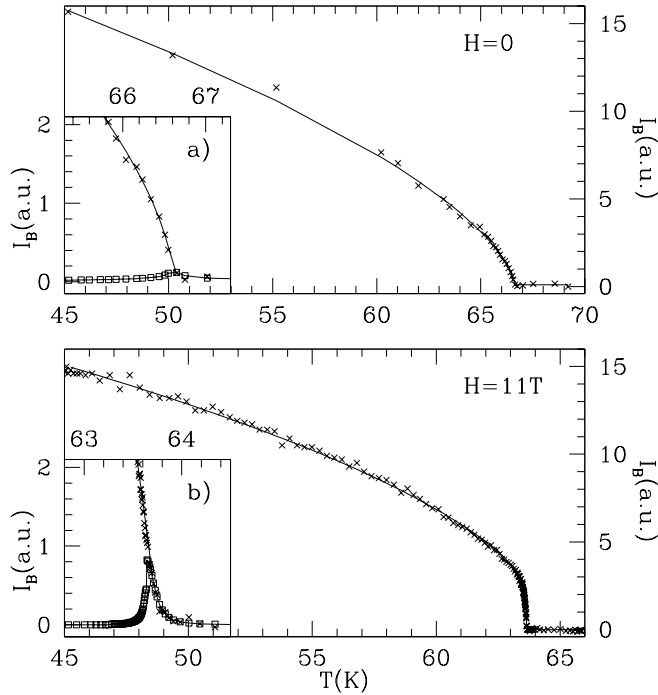}
 \vspace*{3mm}
 }
\caption[Bragg Scattering Intensity vs.\ T at $H=0$ and $H=11$~T]
{ The Bragg ZFC intensity, $I_{\rm B}$, versus $T$ for a) $H=0$
and b) $H=11$~T
with the momentum and temperature independent background
subtracted.  The results
in the insets show the critical
scattering contributions to the x-ray
intensity for $H=0$ and $H=11$~T, respectively, determined from
neutron scattering measurements, as described in the text.}
\label{fig:rawdata}
\end{figure}

\section{QUASI-STATIONARY CRITICAL BEHAVIOR}

Figure \ref{fig:rawdata} shows the $(1 \ 0 \ 0)$ Bragg intensity at
$H=0$ and, for ZFC,
at $H=11$~T versus temperature, with the momentum and temperature
independent background subtracted.  The background depends on the
precise experimental configuration, but not on the
thermal cycling procedure used in collecting data.  The background
is mostly from sources other than the crystal itself.
For comparison of the background
to the Bragg intensity signal, typical background
counts for the $H=11$~T scans were
eight counts per second whereas the $q=0$
intensity was 300 counts per second at $T=37$~K.
Above the transition, the background-subtracted intensity
at small $q$ results
only from the critical scattering and goes
to zero well above $T_{\rm c}(H)$, indicating that there are
no discernible contributions from multiple scattering to the
measured Bragg intensities.  To determine the
critical scattering for the $H=0$ and $11$~T scans,
neutron scattering line shapes, obtained with a sample of
nearly the same magnetic concentration~\cite{ymkybsfa03} using
a procedure described previously~\cite{yzllbgl02,sbf99},
were folded with the x-ray resolution, and the overall $q=0$ amplitude
was adjusted to fit the $H=11$~T data above $T_{\rm c}(H)$.
Insets a and b in Fig.\ \ref{fig:rawdata} show the critical scattering
contributions for $H=0$ and $11$~T, respectively.
As a result of the high momentum resolution of the x-ray technique,
the critical scattering contributions at $H=0$, which are
nearly Lorentzian for $H=0$~\cite{sbf99,bybf04}, are almost negligible
(inset a).  For $H>0$ (inset b), however, the critical scattering
has been shown to be non-Lorentzian and to have a much
larger intensity at small $q$~\cite{sbf99}.  Recently, it
has been shown that this scattering is consistent with
fractal spanning clusters, nucleated at random-field
pinning sites, that form and grow as as the transition temperature is
approached from above~\cite{ymkybsfa03}. 
Consequently, a small contribution to the $q=0$ scattering
is more discernible for the $H=11$~T data.  After the critical
scattering contribution is subtracted from the overall Bragg
scattering intensity, the order parameter exponent can be
determined using Eq.\ \ref{eqn:xray-order}.
As shown previously~\cite{yzllbgl02}, the exponent determination is
not very sensitive to the details of the background
subtraction.

Although neutron scattering measurements
using $\rm Fe_{0.85}Zn_{0.15}F_2$~\cite{ymkybsfa03}
and $\rm Fe_{0.93}Zn_{0.07}F_2$~\cite{sbf99} show no evidence
for micro-domain structure formation in the critical scattering,
$H>0$ hysteresis in $I_{\rm B}$ is evident.
FC intensities are larger than the ZFC ones, which is a result
of extinction.  For ZFC samples, the ordering is so
perfect that relatively few neutrons satisfy the Bragg condition.  With FC,
even with no large-scale
domain structure, there is apparently enough disorder
to strain the crystal, through magnetostrictive
effects, allowing more neutrons to
scatter.  The x-ray Bragg scattering also
shows hysteresis, but in this extinction-free case
the ZFC data are higher in intensity.
FH data are intermediate between
the ZFC and FC curves.
We note that specific heat critical behavior
measurements also show hysteresis
very close to $T_{\rm c}(H)$ at this concentration~\cite{ysmb04,sb98}.

The Bragg intensity curves for $H>0$
in Fig.\ \ref{fig:rawdata} clearly
approach $T_{\rm c}(H)$ {\em vertically}.  This is characteristic
of experiments~\cite{ymkybsfa03,sbf99,bybf04} and
simulations~\cite{bb00a,pb05} for $x>x_{\rm v}$ and is in stark contrast with
experiments~\cite{b00,hfbt93,bwshnlrl96} and simulations~\cite{bb00b}
for $x<x_{\rm v}$,
where $I_{\rm B}$ approaches $T_{\rm c}(H)$ {\em horizontally}.
The latter behavior
is attributable to micro-domain structure formation, which is
energetically favorable when the vacancies percolate through
the crystal, as shown in Monte Carlo simulations~\cite{bb00a}.

\begin{figure}[t!]
 \centerline{
  \epsfxsize=3.8in
   \epsfbox{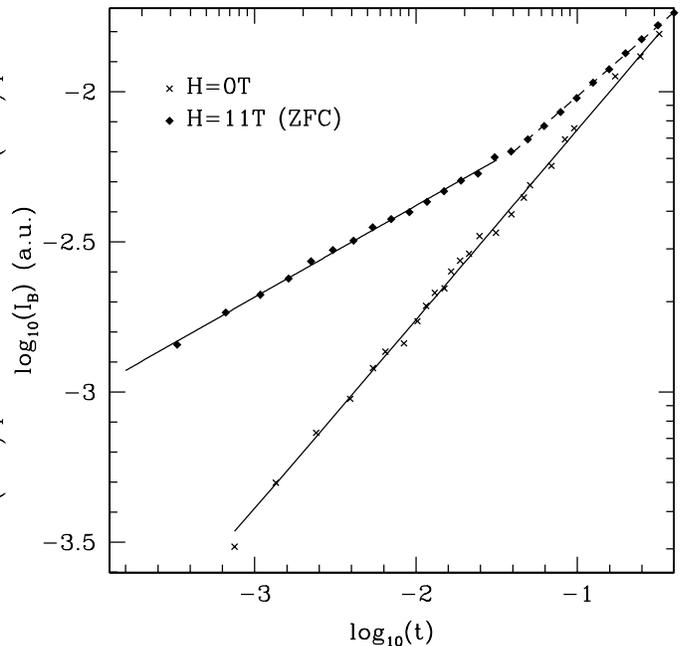}
 \vspace*{3mm}
 }
\caption[Logarithm of Bragg Scattering Intensity vs.\
$\log_{10}{t}$ at $H=0$ and $H=11$~T]
{The same ZFC data as in Fig.\ \ref{fig:rawdata}, corrected for
the critical scattering contribution, plotted as
the logarithm of the intensity versus the logarithm of $t$.
The solid line for $H=11$~T indicates RFIM behavior
with $\beta = 0.17$, while the solid line for $H=0$
reflects conventional random-exchange behavior with $\beta = 0.35$.
}
\label{fig:rawdatafit_11t}
\end{figure}

Figure \ref{fig:rawdatafit_11t} shows the logarithm of $I_{\rm B}$, with the constant
background and critical scattering contributions subtracted,
vs.\ the logarithm of $t$ for $H=0$ and for $11$~T under ZFC cycling.
The values of $T_{\rm c}(H)$ were determined from fits to the data.
For $0.0007<t<0.03$ and $H=0$, we find $\beta = 0.35 \pm 0.02$
(lower solid line), which agrees well with
several experimental and theoretical determinations for the random-exchange
Ising model, as recently discussed~\cite{b00,pv02,fhy03}.
For $11$~T, a crossover from random-exchange to RFIM critical behavior occurs
near $t=0.03$, consistent with
birefringence measurements~\cite{ysmb04},
and the data can be fit to a single power law only
in the range $0.0001<t<0.03$.  The fit over this range
yields the exponent $\beta = 0.17 \pm 0.01$
for $H=11$~T, as
indicated by the upper solid line in Fig.\ \ref{fig:rawdatafit_11t}.

\begin{figure}[t!]
 \centerline{
  \epsfxsize=3.8in
   \epsfbox{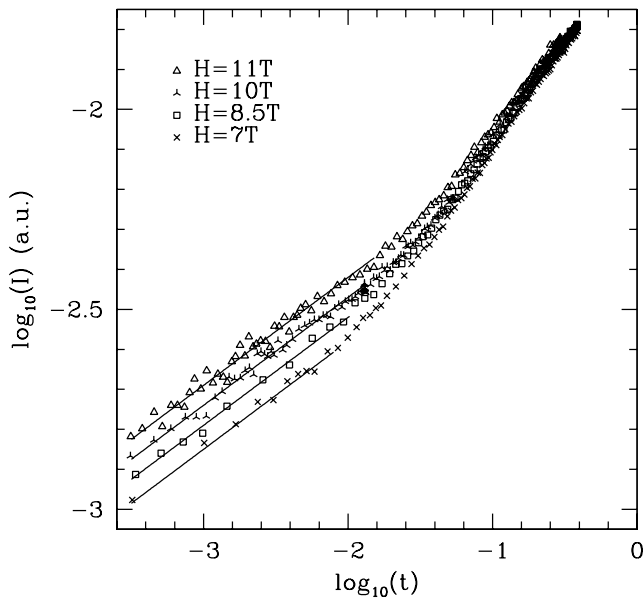}
 }
\caption[Logarithm of Bragg Scattering Intensity vs.\ $\log_{10}{t}$ in
Different Fields]
{The logarithm of Bragg peak intensities $I_{\rm B}$ vs.\ temperature in
fields of $7$, $8.5$, $10$, and $11$~T.
All data have been corrected for the critical scattering
contribution.  Solid lines with a slope of $2\beta=0.34$
are added to the data set at each field. The amplitudes of the data
sets are adjusted to agree at large $t$.
\label{fig:logplot}
}
\end{figure}

In addition to the $H=11$~T data described above, we
also measured the order parameter
at $H=7$, $8.5$, and $10$~T to investigate
crossover effects.
All of the data were taken upon implementing the ZFC protocol.
After we realigned the $(1 \ 0 \ 0)$ Bragg peak
position at $H=11$~T using transverse scans,
the sample was slowly warmed through the transition while the intensity,
$I_{\rm B}$, was recorded.
The rate of temperature change was controlled to be $0.2$~K/min away from
the transition, and was decreased to $0.02~$K/min close to the transition.
After the sample was warmed above the phase
boundary and subsequently cooled, we re-examined the beam alignment to
ensure that the peak intensity was measured.
We found the peak position to change very little upon cycling.
The same procedures were repeated for fields of $10$,
$8.5$, and $7$~T. 
Figure \ref{fig:logplot} summarizes the ZFC order
parameter measurements at various fields.
Three common features can be discerned.
First, far below the transition temperature, the Bragg peak
intensity tends toward saturation since all
magnetic moments in the beam-illuminated
region achieve long-range order at such fairly low temperatures.
Second, for reduced temperatures in the range
$10^{-1.5}<t<10^{-0.5}$,
the behavior of $\log_{10}(I)$ is very
similar to the REIM behavior, as shown in Fig.\
\ref{fig:rawdatafit_11t}; the difference in the
slope results from the variation of $T_{\rm c}$ with field,
since the transition temperature is determined by
the random-field behavior.
Third, as the temperature
approaches $T_{\rm c}(H)$, the logarithm of $I_{\rm B}$ in different fields
exhibits the same exponent which is indicated by the parallel
straight lines through the experimental data.
The fitted exponents of the order parameter at $H=7$, $8.5$, $10$ and
$11$~T are $\beta = 0.18\pm 0.02$, $0.18\pm 0.02$, $0.17\pm 0.02$ and
$0.17\pm 0.01$, respectively, where the quoted errors are
statistical.  The higher fields provide larger
ranges of reduced temperatures over which to fit the data and
so yield more reliable values for $\beta$.  From these results, we obtain
a more precise order-parameter critical
exponent, $\beta = 0.17 \pm 0.01$, than reported
in our previous study~\cite{yzllbgl02}.

\begin{figure}[t!]
 \centerline{
  \epsfxsize=3.8in
   \epsfbox{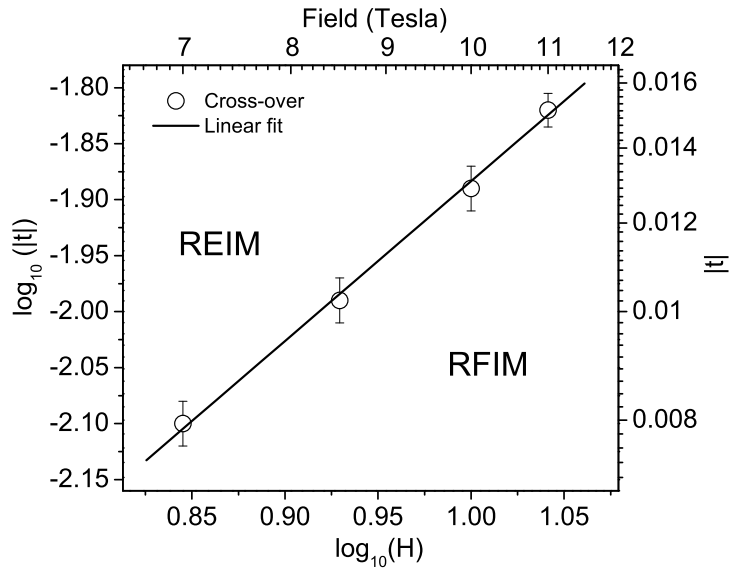}
 }
\caption[Crossover from REIM to RFIM]
{ The crossover points from REIM to RFIM critical behavior for several
fields. The solid line has a slope of $1.42$.
}
\label{fig:crossover_exp}
\end{figure}

In Fig.\ \ref{fig:crossover_exp} is
plotted the logarithm of $H$ versus the logarithm of $t$
where REIM crosses over to RFIM critical behavior.
This value is defined, for the data shown in Fig.\ \ref{fig:logplot}, as the
intersection of a straight line through asymptotic RFIM behavior
and one through the REIM behavior.
From the scaling variable $tH^{-2/\phi}$,
we see that the slope in Fig.\ \ref{fig:crossover_exp}
yields the random-exchange to random-field
crossover exponent $\phi$.  Indeed, the value obtained, $\phi=1.4 \pm 0.05$,
agrees well with the earlier experimentally determined and the
most recent theoretically established values, both of which are
$\phi = 1.42 \pm 0.02$ for three
dimensions~\cite{b00,cpv03,kjbr85,mkj86,kkj86}.

Through the Rushbrooke scaling relation
\begin{equation}
2\beta + \gamma + \alpha \ge 2 \quad ,
\end{equation}
which is usually satisfied as an equality, $\beta$ is related to the
universal critical exponents $\alpha$ (for the specific heat) and
$\gamma$ (for the staggered susceptibility) of the $d=3$ RFIM.  The
experimentally determined specific heat peak is nearly logarithmic
and very symmetric close to $T_{\rm c}(H)$, consistent with $\alpha
= 0.00 \pm 0.01$~\cite{ysmb04,sb98}.  Recent neutron scattering
analyses~\cite{ymkybsfa03,sbf99} yield a value of $\gamma = 1.68\pm 0.03$,
consistent with earlier results.
Therefore, the experimental value $\beta \approx 0.17 \pm 0.01$ is
consistent with Rushbrooke scaling
$2\beta+\gamma+\alpha=2.02\pm 0.06$, even though the system is
clearly not in equilibrium.

None of the ZFC results are dependent on the rate at which
the sample was warmed, typically between $0.2$~K/min and
$0.02$~K/min.   However, the results are quite
sensitive to temperature reversals of even a few
mk, including overshoots of the set point when stabilizing
the temperature.  Such overshoots were
meticulously avoided in the critical behavior measurements.

\section{THERMAL HYSTERESIS}

We next turn our attention to the effects of hysteresis and
temperature reversals, including the difference between
the critical behavior observed upon ZFC and FC.  The order parameter
measurements display significant irreversibility in both $T$ and $H$
cycling procedures.  We will address field hysteresis effects
in the next section. 

\begin{figure}[t!]
 \centerline{
  \epsfxsize=3.8in
   \epsfbox{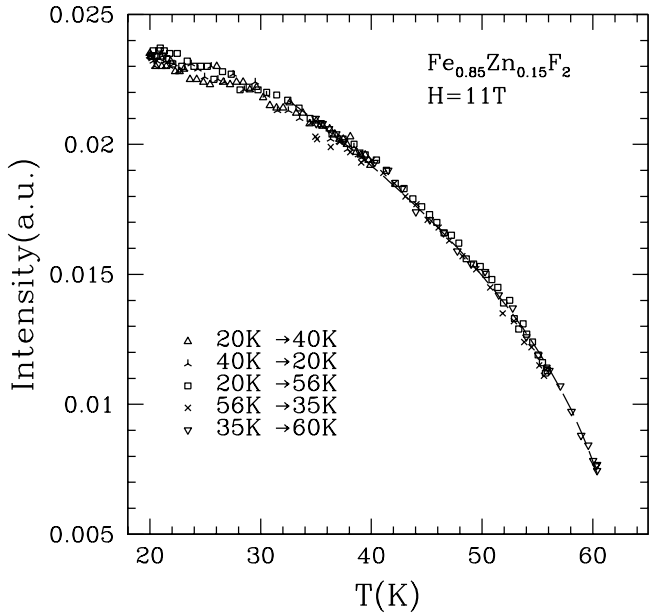}
 }
\caption[Bragg Scattering Intensities vs.\ T in the Range of
$20$~K and $60$~K]
{The temperature dependence of the Bragg peak intensities
measured in the temperature range of $20$~K and $60$~K with
$H=11$~T. The dashed line is a guide to the eye.
}
\label{fig:11tcycle_low}
\end{figure}

A ZFC temperature cycle, followed by repeated cooling
and heating at $H=11$~T, is shown in
Fig.\ \ref{fig:11tcycle_low} for the
temperature range between $20$~K and $60$~K. There is no observable
hysteresis in this temperature range; the data taken upon cooling
are essentially identical to those obtained upon heating.

\begin{figure}[t!]
 \centerline{
  \epsfxsize=3.5in
   \epsfbox{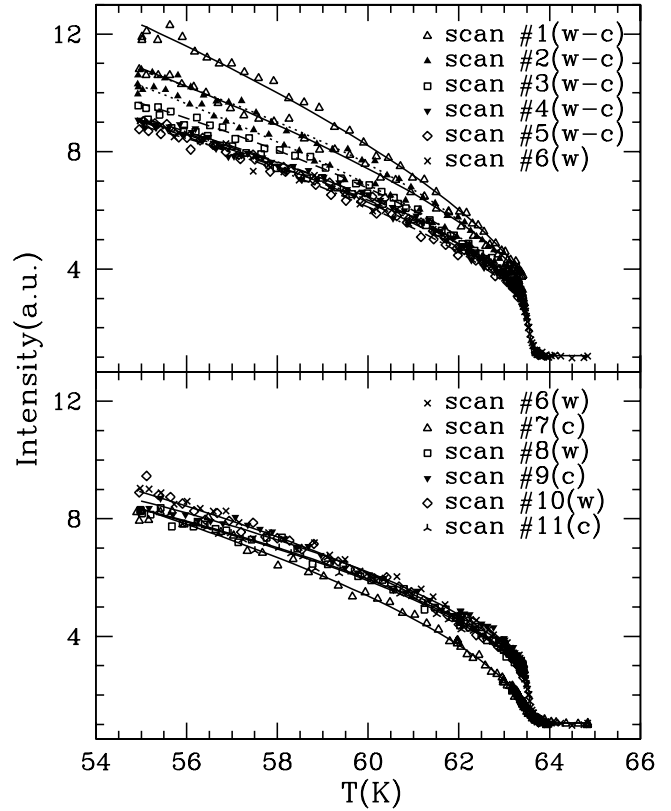}
 }
\caption[Temperature Reversal Curves Below the Transition]
{Behavior of the Bragg scattering upon temperature
reversals after ZFC 
for $H=11$~T (top panel), and after the sample was
FC to $40$~K (bottom panel). In the upper panel, ZFC Bragg peak
intensities at $H=11$~T, as well as after several reversals in
temperature below $T_{\rm c} = 63.7$~K, are shown.
The data were obtained in pairs of warming and cooling
scans (w-c), except for the final set, which was obtained only
upon warming (w).
The curves are guides to the eye.
In the bottom panel, intensities from the final warming through $T_{\rm c}$
shown (curve \#6) in the previous figure are plotted along with FC intensities
and reversal intensities at $H=11$~T.  The data shown were obtained
in warming (w) or cooling (c) scans.  The curves are guides to the eye.
}
\label{fig:revhys_compare_2}
\end{figure}

However, if we heat the sample to a temperature sufficiently close to
the phase boundary, then, upon reversing the temperature,
the intensity of Bragg scattering shows significant irreversibility, as
shown in the upper panel of
Fig.\ \ref{fig:revhys_compare_2}. The data obtained
upon cooling exhibit lower intensities compared to the
data obtained upon the initial ZFC warming.  A significant discrepancy
is observed after the sample is cooled to $T=55$~K.  The behavior of temperature
reversal amplitudes of the order parameter after cooling with
$H=0$ and raising the field was further
investigated by repeatedly cooling and warming the sample, using
rates described earlier.
The top panel shows the measurements taken upon cycling
after cooling with $H=0$ and raising the field without exceeding
$T_{\rm c}$ at $H=11$~T.
All the reversals close to $T_{\rm c}$ take place at $T=63.40$~K,
which is $300$~mK below $T_{\rm c}$. For reversal measurements
without crossing the phase boundary, the amplitudes of the Bragg
peaks gradually decrease after each cycle, but the warming curve follows
the previous cooling curve quite well. In this way, the overall amplitudes
slowly decrease upon cycling. After several cooling and warming loops, the
amplitude is finally stabilized at the lowest cooling curve (scan \#5)
shown in the top panel.  The last set of data (scan \#6)
was obtained by warming through the phase boundary.

In the lower panel of Fig.\ \ref{fig:revhys_compare_2},
the data obtained while warming across the
phase boundary are repeated to serve as a reference.  The sample was
cooled using the FC procedure to $55.00$~K.
The same temperature loop measurements were then repeatedly taken
between $55.00$~K and $63.40$~K. Contrary to the cycling shown
in the upper panel, the cooling and warming curves do not exhibit
as clearly a slow approach to the stabilized equilibrium state from
above, but
instead seem to reach the stabilized behavior quickly from
below upon warming.
Further thermal cycling does not change the intensity.
The initial FC curve is quite distinct from all other heating and
cooling curves below $T_{\rm c}$.  The FC transition appears to be
at the same temperature as the ZFC one, within experimental accuracy.

\begin{figure}[t!]
 \centerline{
  \epsfxsize=3.8in
   \epsfbox{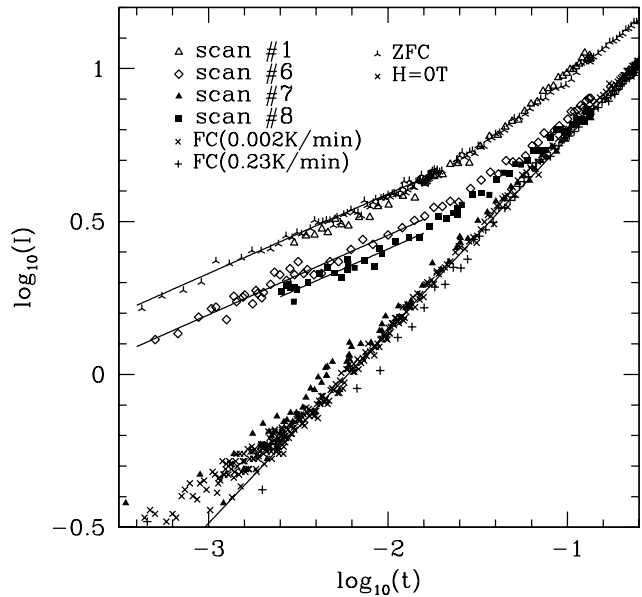}
 }
\caption[Logarithm of Scattering Intensities upon Reversal]
{A subset of the same data shown in Fig.\ \ref{fig:revhys_compare_2},
plotted as the logarithm of the amplitudes versus the logarithm of $t$.
Also included are data from a scan at $H=0$ as well as a fast
FC scan.  The fast scan was obtained by cooling through the transition
at a rate of $0.23$~K per minute. The data from scan \#1 are only
those obtained upon warming.
The top two curves represent ZFC and the next two are
FC and heating after FC.  They all indicate a critical
exponent consistent with $\beta=0.17\pm 0.01$.
The solid line at the bottom has a slope indicating $\beta = 0.35$.
}
\label{fig:xray-fcfhlog}
\end{figure}

In Figure \ref{fig:xray-fcfhlog}, the logarithm of the peak amplitude
versus logarithm of $t$ is plotted for various heating and
cooling procedures. Only data of some typical
scans are shown. Clearly, ZFC and FH curves for $H=11$~T have the
same critical power-law behavior with $\beta=0.17$ at small $t$.
For temperatures outside the random-field critical
region, the data cross over from the RFIM asymptotic critical
behavior to the random-exchange Ising behavior as $t$
increases.  The order parameter
measured using the FC procedure shows strikingly
different behavior from that of
ZFC and FH ones; the overall intensities stay much lower and there
is no observable crossover from REIM to the RFIM
behavior. The exponent for the
FC has a value of $\beta \approx 0.35$, but perhaps shows
some rounding at $t$ smaller
than $0.002$.  Over a large range of reduced temperature, the results
obtained upon FC seem very similar to the
random-exchange Ising model. Two different cooling rates,
$0.002$ and $0.23$~K per minute, were used for the FC protocol.
Even though these rates differ by two orders of magnitude,
we find that the measured intensity is essentially insensitive to the
choice of cooling rate.
In order to compare the $H=0$ and FC data, the zero-field peak
intensity is multiplied by $0.70$, which reflects the difference
of line shapes for the two cases, as we will discuss below.

\begin{figure}[t!]
 \centerline{
  \epsfxsize=3.8in
   \epsfbox{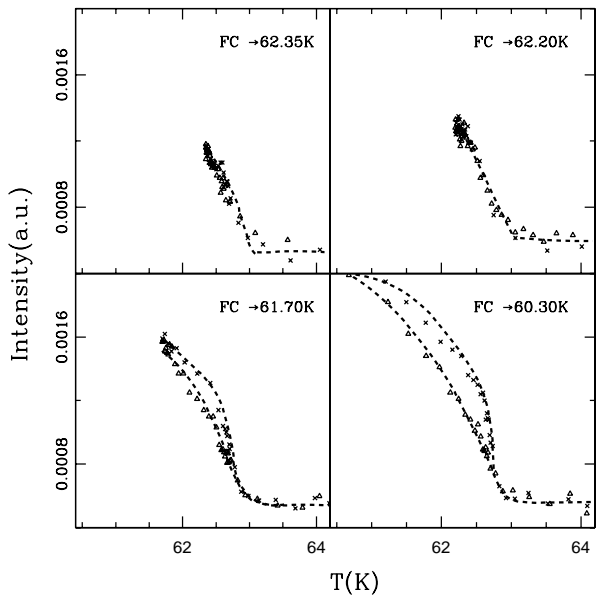}
 }
\caption[Loop measurements for FC at different Temperatures]
{Cycling measurements at $H=10$~T after FC with low-temperature reversals at
four different temperatures, $62.35$, $62.20$, $61.70$
and $60.30$~K. The warming curves
gradually exhibit normal FH features as the reversal temperature
moves away from $T_{\rm c}$, but the cooling curve is always lower. The
triangles represent data under FC conditions and the
crosses represent data under warming conditions.
}
\label{fig:xray-fchys}
\end{figure}

Thermal cycling loops at $H=10$~T after FC are
shown in Fig.\ \ref{fig:xray-fchys}.
As the reversal temperature is lowered away from
$T_{\rm c}(H)$, the warming curves, above the first
few points after the reversal, gradually display normal FH behavior.
For data taken with the lower temperature reversal point very close
to $T_{\rm c}$, there is little
discernible difference between warming and cooling data.

\begin{figure}[t!]
 \centerline{
  \epsfxsize=3.8in
   \epsfbox{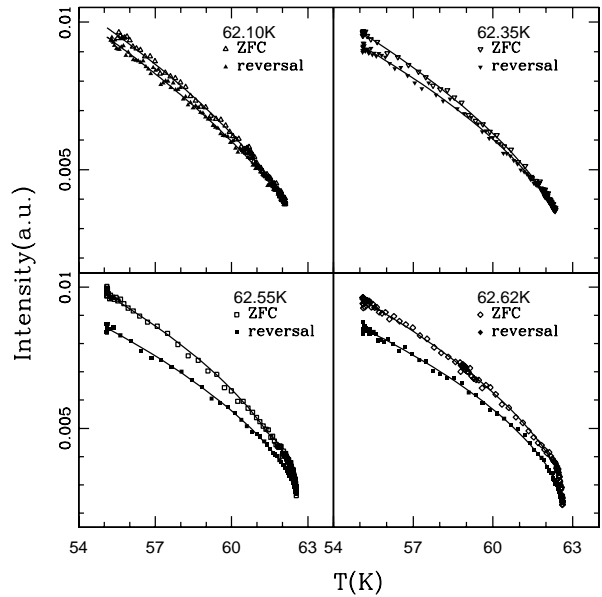}
 }
\caption[Reversal Curves near the Phase Boundary at $H=10$~T]
{A series of reversal curves near the phase boundary at
$H=10$~T.  The reversal points are close to, but below
the phase boundary.}
\label{fig:reversehys}
\end{figure}

The irreversibilities of the random-field Ising model order parameter
are also reflected in the peak intensity difference between ZFC and reversal
curves at low temperature.  The size of the intensity difference depends
on how close one chooses to reverse the temperature near the phase boundary,
as shown in Fig.\ \ref{fig:reversehys}.  Four points close to
$T_{\rm c}=62.85$~K at $H=10$~T were
chosen to reverse the temperature: $62.10$~K; $62.35$~K;
$62.55$~K; and $62.62$~K.  Each reversal measurement was taken after
ZFC preparation.  The intensities for ZFC curves are
normalized at the lowest temperature so that they all have the
same amplitude for comparison.
For reversals close to $T_{\rm c}(H)$, 
the shapes of the reversal curves in the critical region are quite
different from the ZFC ones.  The shapes for the cooling curves
suggest a larger $\beta$ since they appear not to be as steep
as the ZFC ones.

\begin{figure}[t!]
 \centerline{
  \epsfxsize=3.8in
   \epsfbox{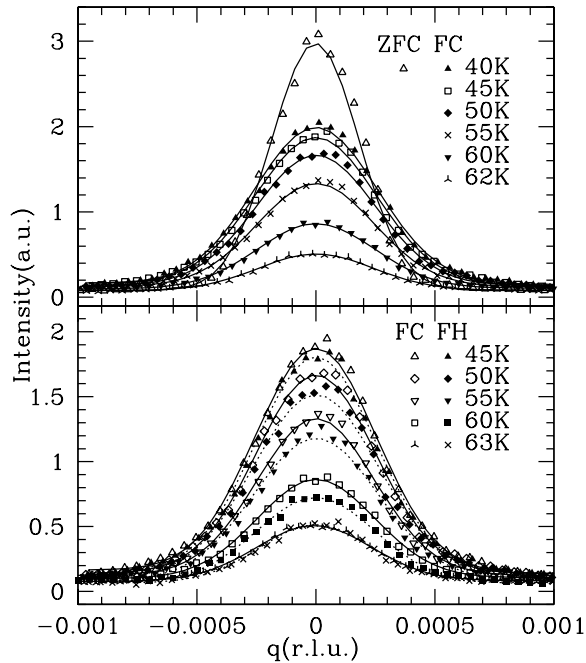}
 }
\caption[Transverse Scans upon FC at $H=10$~T]
{Representative transverse x-ray scattering scans for
the FC and FH protocols. Data were taken at $H=10$~T. The solid curves are
the results of least squares fits to a Gaussian shape. A scan taken
under ZFC conditions is also plotted in the upper panel for comparison.
}
\label{fig:comparefcfh}
\end{figure}

\begin{figure}[t!]
 \centerline{
  \epsfxsize=3.8in
   \epsfbox{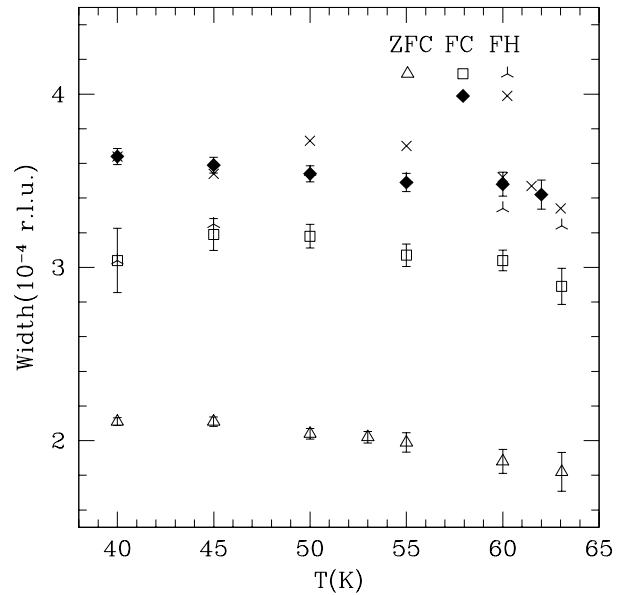}
 }
\caption[HWHM from Gaussian fits for x-ray Scattering Line Shapes
for ZFC, FC and FH]
{The HWHM of x-ray scattering line shapes for ZFC, FC and
FH.  ZFC line shapes are much narrower than those for FC and FH. In the
graph, representative error bars are shown.}
\label{fig:width}
\end{figure}

The temperature dependences of the x-ray scattering line shapes are
shown in Fig.\ \ref{fig:comparefcfh}
under zero field cooled, field cooled
and field heated conditions for $H=10$~T.
Under ZFC, the data
have typical Gaussian shapes with a half-width-at-half-maximum
(HWHM) of $=2.1\times 10^{-4}$~r.l.u.  For FC and
FH conditions, the line shapes and intensities are distinctly different
from ZFC.
The tails in the former cases are much larger and the central intensities
are smaller.  The temperature dependence of the widths for ZFC, FC and
FH is shown in Fig.\ \ref{fig:width}.  All data were fit using Gaussian line
shapes. The HWHM from Gaussian fits for the FC and FH line shapes
is much larger than that for ZFC.
Gaussian fits describe the data well and
indicate $3.5\times 10^{-4}$~r.l.u. for the widths under FC and FH.
Although it is not known if Gaussian fits are the correct ones
to use for FC and FH, they do work well and afford direct comparisons
of widths for the three cases.

\begin{figure}[t!]
 \centerline{
  \epsfxsize=3.8in
   \epsfbox{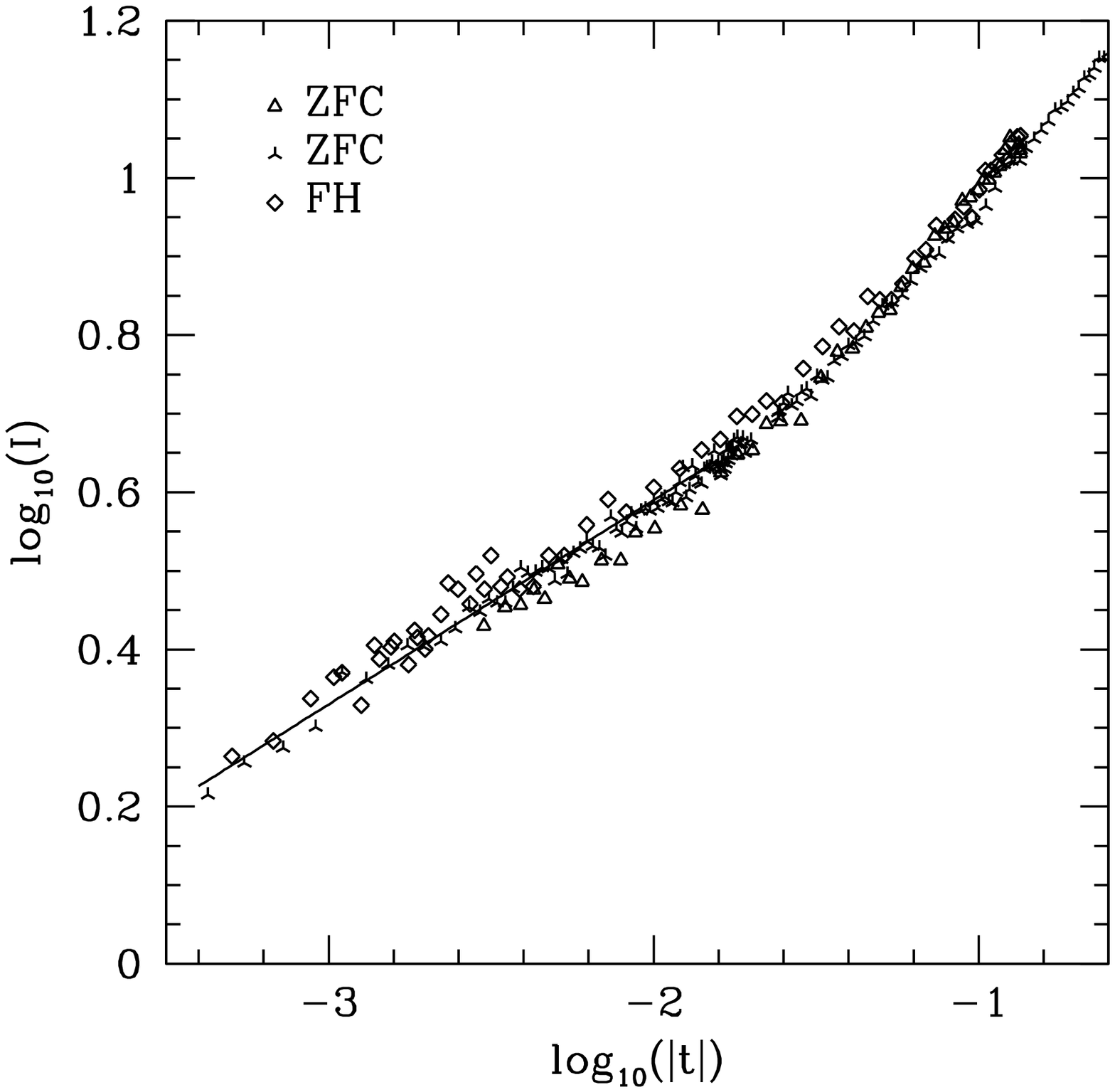}
 }
\caption[Logarithm of Integrated Scattering Intensities upon ZFC
and FH]
{The logarithm of the integrated scattering intensity of
transverse scans vs.\ the logarithm of $t$ for ZFC and FH
protocols.  The two sets of ZFC data represent experiments that
were separated by several months in time.
}
\label{fig:newzfcfh}
\end{figure}

In Fig.\ \ref{fig:xray-fcfhlog}, the peak intensities are plotted
versus reduced temperature. For a more meaningful comparison,
we should compare the integrated intensity versus $t$.
Figure \ref{fig:comparefcfh} shows the shape of transverse scans for ZFC and FH.
In both cases, the line shapes
appear to be Gaussian, but with different peak widths.
If we take that peak broadening into account, the integrated intensity
should be the product of peak width and peak intensity. Figure
\ref{fig:newzfcfh} shows the logarithm of integrated intensity versus
the logarithm of $t$.  Indeed, within the experimental accuracy,
the FH data collapse onto the ZFC data curves.

\begin{figure}[t!]
 \centerline{
  \epsfxsize=3.8in
   \epsfbox{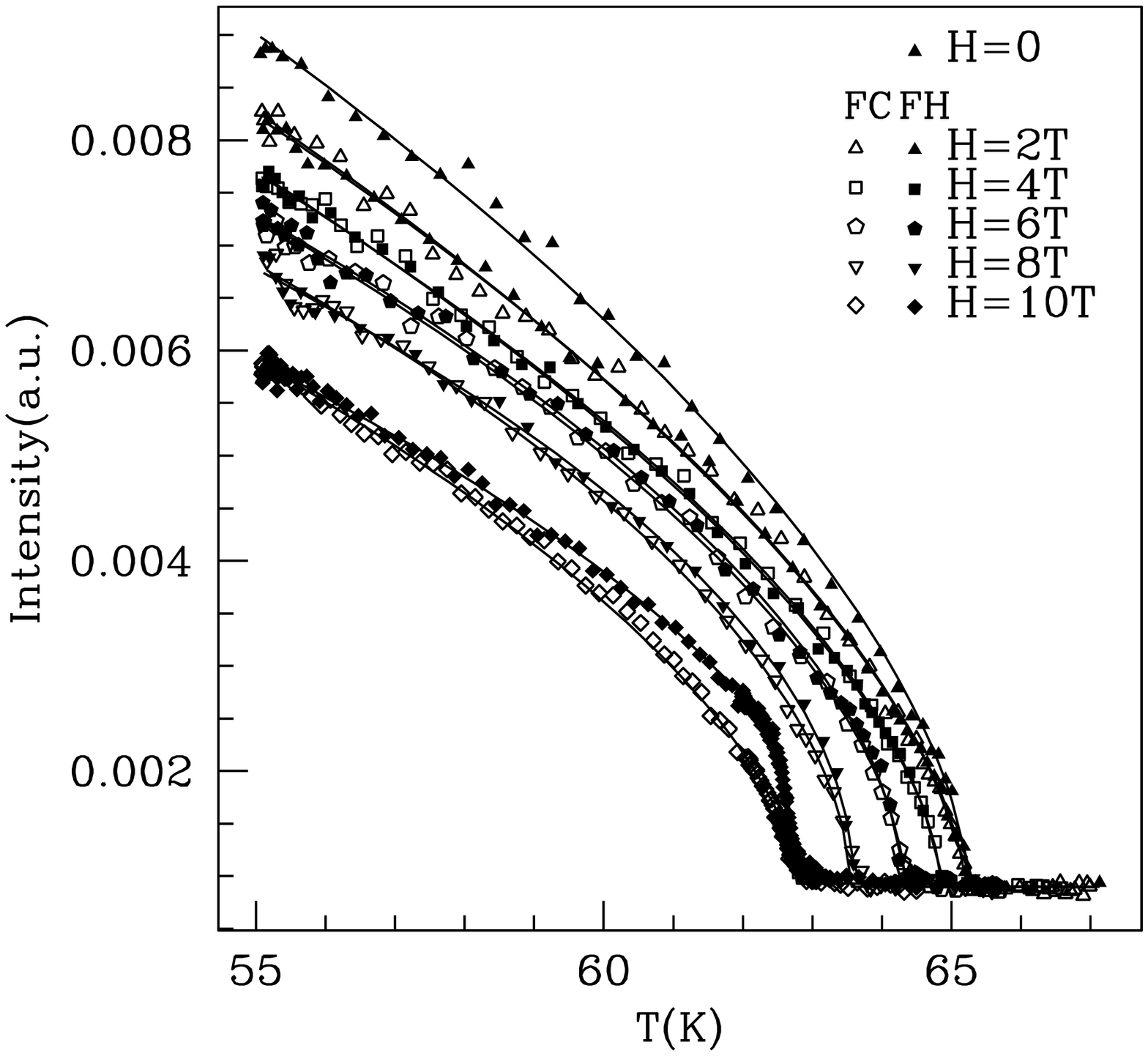}
 }
\caption[Field Dependence of Bragg Scattering]
{Field dependence of the Bragg scattering for
$0\le H \le 10$~T. The difference
between FC and FH increases as the field increases.
The FC intensity at low temperature becomes lower than
the FH intensity.}
\label{fig:fieldhys}
\end{figure}

Figure \ref{fig:fieldhys}
shows the difference between FH and FC data sets increasing with the
strength of the applied field. The hysteresis is difficult to discern
in these measurements for $H<8$~T, whereas
for $H=10$~T the different shapes for FH and FC are quite
evident.

\section{FIELD HYSTERESIS}

\begin{figure}[t!]
 \centerline{
  \epsfxsize=3.8in
   \epsfbox{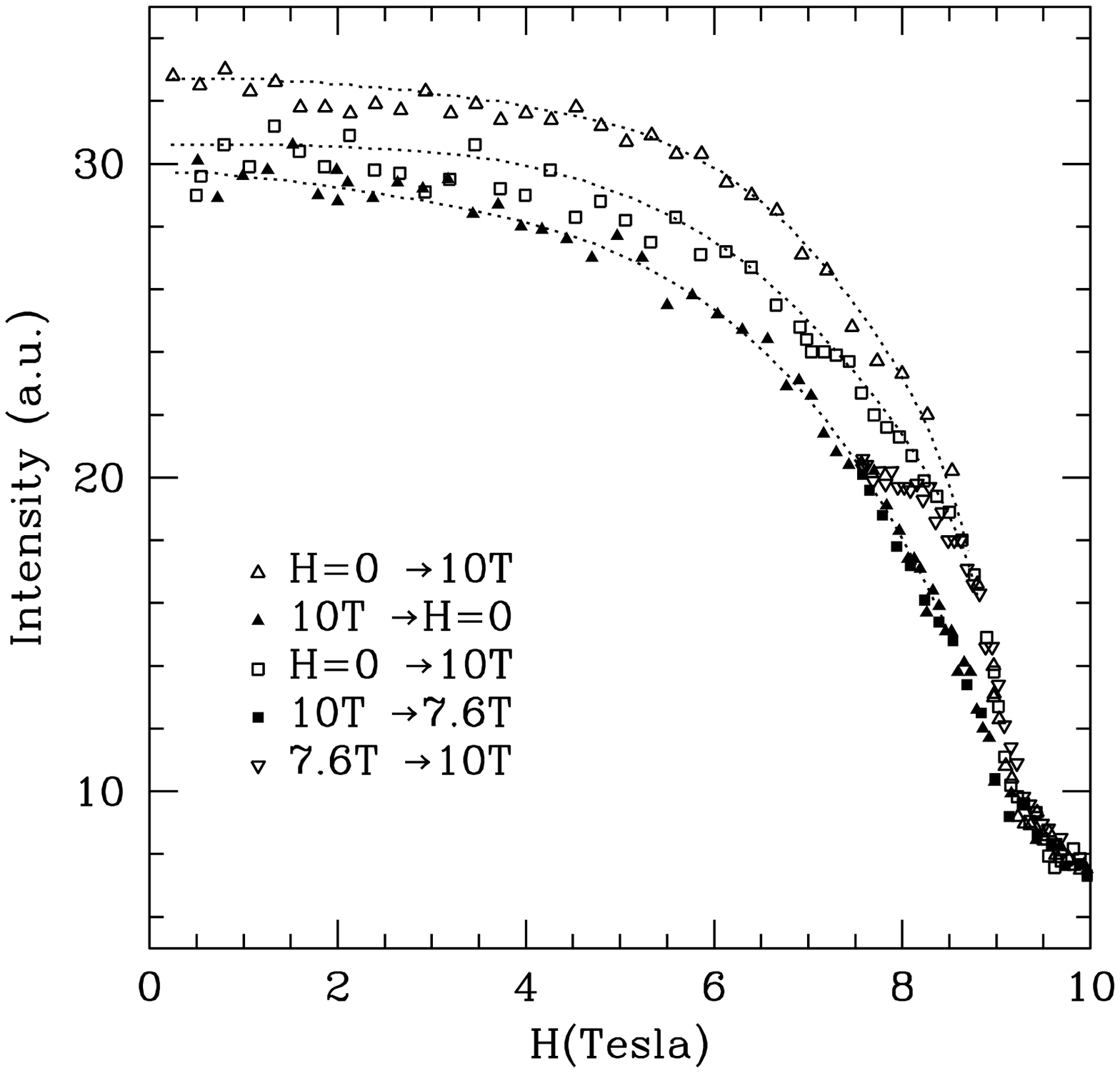}
 }
\caption[Field Cycling Bragg Scattering at $T=63.10$~K]
{Comparison of the field cycle Bragg scattering data at
$T=63.10$~K. Open symbols represent data from `field-raising' and
solid symbols represent data from `field-lowering'.  The dotted
curves are guides to the eye.}
\label{fig:fieldscan_tc}
\end{figure}

Finally, field-cycle measurements were carried out to study
the history dependence near $T=63.10$~K, the transition
temperature for $H=9$~T. The sample was first warmed into
the paramagnetic state at $70$~K, then cooled in zero field to
$63.10$~K. The intensity was recorded as the field was slowly raised
to $11$~T, where $T_{\rm c} = 63.7$~K, followed by a series of field
cycling procedures with the
temperature held at $63.10$~K. The result is presented in
Fig.\ \ref{fig:fieldscan_tc}. The general features of the experimental
data are similar to those of the thermal cycle measurements in
Fig.\ \ref{fig:revhys_compare_2}.  The
initial ``field raising'' data, prepared by cooling in zero field and
represented by open triangles, stay the highest and the intensity
goes to zero rapidly as $T_{\rm c}(H)$ is approached.
As the field is lowered,
significantly less of the peak Bragg intensity is recovered, as shown for
data represented by the solid symbols.
When the field is raised after first lowering it,
the Bragg intensity initially has an intensity similar
to that at low field, but eventually crosses over
to the initial ``field raising'' curve, in the vicinity of $H_{\rm c}$.  The
curvature is more like the initial ``field raising'' procedure near
the transition.  Similar behavior is seen upon raising the field
beginning at $H=7.6$~T.

\section{CONCLUSION}

In summary, we used extinction-free x-ray scattering to
study the hysteresis of the order parameter of random-field
Ising model.  The critical exponent $\beta=0.17\pm 0.01$ is
obtained for ZFC, whereas a different critical-like behavior,
very similar to the random-exchange behavior seen for $H=0$, takes
place for FC.  The crossover from random-field Ising model
to random-exchange Ising model was observed and
the value $\phi=1.40\pm 0.05$ obtained for the crossover
exponent is consistent with the earlier studies and theory.
We examined the history-dependent critical behavior in
detail for various thermal and field cycling.
It is clear from the different cycling experiments that
the system under FC is not in equilibrium.  However, the
critical behavior upon ZFC is rate independent and
is consistent with a second-order transition if the
temperature is never reversed.  The ZFC order-parameter critical
behavior, {\em i.e.} the power law behavior with $\beta = 0.17 \pm 0.01$
and the crossover to REIM behavior,
must be associated with rather stable, quasi-stationary states of
the magnetic long-range component of the order.  FH data
exhibit line shape widths larger than ZFC and similar to FC,
suggesting that some disorder introduced upon FC remains.
Aside from the slightly wider line shapes, the critical behavior
of the FH data closely resembles
the ZFC behavior.  This suggests that it is the heating itself that
is important in the manifestation of the new RFIM critical behavior
and not the history of the temperature-field cycling.

Although these findings are not in good agreement with equilibrium
simulations and ground state calculations, as mentioned earlier, they
are in good agreement with non-equilibrium Monte Carlo
studies~\cite{sybzb04}.  Furthermore, they are consistent with
studies on uniaxial relaxor ferroelectrics, a rather
different experimental realization of the random-field
Ising model~\cite{gwwik04,kdlbzp02}
in which the order-parameter critical exponent is observed to
increase from $\beta =0.13$ to $\beta =0.30$, a value close
to the random-exchange value, when the initial polarization is
varied from 100\% to 0.8\%.  It is argued for the
ferroelectric system that a lower initial
polarization corresponds to a compensation of the random-field
by domain walls.  In the present case of the dilute
antiferromagnet in a uniform field, it is clear that the large suppression
of the transition is still present upon FC and
that random fields are therefore not suppressed.
However, it is also clear that the new critical behavior
that is observed upon ZFC does not occur when the
sample is FC.  Mean-field treatments~\cite{lmp95,d06}, exact ground state
calculations~\cite{mm05} and Monte Carlo simulations~\cite{mf06}
indicate a very complex energy landscape and unusual
characteristics near the RFIM phase transition.
Perhaps in FC the fractal spanning cluster structures
formed in our sample as it is cooled
towards $T_{\rm c}(H)$
influence the character of the ordering process upon FC~\cite{ymkybsfa03,spa02}.
The spanning clusters form as the transition is approached
from above and represent an ordering process quite different
from pure systems.  If the system cannot readily evolve from
that configuration just above $T_{\rm c}(H)$ to long-range
order below, this might indeed result in the severe
hysteresis we have observed.
In the ZFC process, this high magnetic concentration sample starts
from a fully ordered lattice, retaining a single domain structure,
and is perhaps less influenced by spanning cluster structures
that form above $T_{\rm c}(H)$.  Below $T_{\rm c}(H)$, metastability
upon cooling has been described recently in terms of
instantons which are a result of the complicated energy
landscape due to the random fields~\cite{ms06}.
Reversals of the temperature below the transition surely
represent states in between those formed under ZFC and FC.
A theoretical understanding of these differences
and why they occur is lacking at this time.  However, it is
clear that this transition, with its
coexistence of second-order-like critical behavior measured
to very small reduced temperatures as well as
severe hysteresis upon temperature reversals near the
phase boundary, is highly unusual.  It is not correct to assume that simply
not being in equilibrium would account for different critical
behavior~\cite{cadmz04}.  It is likely that such behavior is more
generic in systems undergoing phase transitions in the
presence of quenched disorder.  Although $\rm Fe_xZn_{1-x}F_2$ and
its isomorphs in applied fields are the most characterized examples
of the RFIM, other magnetic~\cite{dglcb04,mgkbgnu00} and ferroelectric~\cite{kdlbzp02,k02}
systems, as well as manganites~\cite{dm01,mmfyd00} have been studied. 

SSRL is operated by Stanford University for the U.S. Department of Energy,
Office of Basic Energy Sciences.
The work at UCSC was funded by Department of Energy Grant
No.\ DE-FG02-05ER46181.  The work at Stanford was supported by the
U.S. Department of Energy under Contract Nos. DE-FG03-99ER45773 and
DE-AC03-76SF00515, by NSF Grant Nos. DMR-9400372 and DMR-9802737, and
by the A.P. Sloan Foundation.  We acknowledge Onuttom Narayan for
useful discussions.

\end{document}